\documentclass[journal]{IEEEtran}

\usepackage{xcolor,soul,framed} %,caption

\usepackage[pdftex]{graphicx}
\graphicspath{{../pdf/}{../jpeg/}}
\DeclareGraphicsExtensions{.pdf,.jpeg,.png}
\usepackage[cmex10]{amsmath}
\usepackage{graphicx}
\usepackage{multirow}
\usepackage{multicol}
\usepackage{amssymb} 
\usepackage{amsmath}
\usepackage{mathtools}
\usepackage{cite}
\usepackage{siunitx}
\usepackage{cuted}
\usepackage{flushend}
\usepackage{multibib}
\usepackage{relsize}
\usepackage{stix}
\usepackage{balance}
\usepackage{array, makecell}

\begin{document}
\bstctlcite{IEEEexample:BSTcontrol}
    \title{High Conversion Efficiency in Multi-mode Gas-filled Hollow-core Fiber}
 \author{ Md. Selim Habib,~\IEEEmembership{Senior Member,~IEEE,} ~\IEEEmembership{Member, OSA,}
      Christos Markos,~\IEEEmembership{Member,~OSA,} 
      
      and Rodrigo Amezcua-Correa,~\IEEEmembership{Member,~OSA} % <-this % stops a space

  \thanks{Manuscript received August XX, 2021.}
   \thanks{M. Selim Habib is with the Department of Electrical and Computer Engineering, Florida Polytechnic University, FL-33805, USA (e-mail: mhabib@floridapoly.edu).}
  \thanks{C. Markos is with the Department of Photonics Engineering, Technical University of Denmark, DK-2800, Denmark (e-mail: chmar@fotonik.dtu.dk).}%
  \thanks{R. Amezcua-Correa is with CREOL, University of Central Florida, FL-32826, USA (e-mail: r.amezcua@creol.ucf.edu).}%
 % <-this % stops a space
  }

% The paper headers
%\markboth{IEEE Photonic Technolgoy Letters, VOL.~XXX, NO.~XXX, June~2021}
%{Habib \MakeLowercase{\textit{et al.}}:XXX}
% title
\maketitle

% Abstract

\begin{abstract}
%\boldmath
 In this letter, an energetic and highly efficient dispersive wave (DW) generation at 200 nm has been numerically demonstrated by selectively exciting LP$_{02}$-like mode in a 10 bar Ar-filled hollow-core anti-resonant fiber pumping in the anomalous dispersion regime at 1030 nm  with pulses of 30 fs duration and 7 $\mu$J energy. Our calculations indicate high conversion efficiency of $>$35\% (2.5 $\mu$J) after propagating 3.6 cm fiber length which is due to the strong shock effect and plasma induced blue-shifted soliton. It is observed that the efficiency of fundamental LP$_{01}$-mode is about 15\% which is much smaller than  LP$_{02}$-like mode and also emitted at longer wavelength of 270 nm. 
\end{abstract}

\begin{IEEEkeywords}
Hollow-core anti-resonant fiber, ultrafast nonlinear dynamics, higher-order mode, dispersive wave.
\end{IEEEkeywords}

\IEEEpeerreviewmaketitle

% I. INTRODUCTION 
\section{Introduction}
%\IEEEPARstart{T}{he} 
\IEEEPARstart{T}{he} ultraviolet (UV) spectral range (100-400 nm) is of great interest in the scientific community because of wide range of applications such as spectroscopy \cite{reinert2005photoemission}, control of chemical reactions \cite{asplund2002ultrafast}, biomedicine \cite{steiner2008medical}, and femtosecond (fs) pump–probe measurements \cite{hockett2011time}. Despite of numerous applications, the availability of such UV sources in this spectral regime are limited and a very few laser sources directly emit UV light and some require complex set-up \cite{kottig2017generation}. These UV laser sources include excimer lasers \cite{ewing1978rare}, cerium fluoride lasers \cite{granados2009mode}, harmonic generation \cite{rothhardt2016100}, and diode
lasers \cite{zhang2019271}. Recently, hollow-core anti-resonant fibers (HC-ARFs) offer many appealing features including broadband guidance, low-light glass overlap, low propagation loss, and weak anomalous dispersion to study ultrafast nonlinear dynamics when filled with noble gases \cite{travers2011ultrafast,joly2011bright,adamu2019deep,adamu2020noise,markos2017hybrid,habib2017soliton}. One of the remarkable features of adding gas in HC-ARF is that both the dispersion landscape and nonlinearity can be tuned by simply changing the pressure of the gas. Based on these properties, several impressive results have been reported % \cite{russell2014hollow,mak2013tunable}. 
to generate light in the UV \cite{adamu2019deep,adamu2020noise,joly2011bright,kottig2017generation}. However most of the works on the UV light generation have been performed considering the excitation of a pure LP$_{01}$-like mode and it is relied on a well-known process called soliton self-compression to the sub single-cycle regime, followed by resonant DW radiation \cite{travers2011ultrafast}. For example, bright emission of UV light was first theoretically predicted in an Ar-filled HC-ARF with efficiency of 20\% \cite{im2010high} and later experimentally reported with efficiency of 6\% \cite{joly2011bright}. Recently, an energetic DW generation at 275 nm is experimentally demonstrated by pumping in the mid-IR regime of 2460 nm \cite{adamu2019deep}.

In this work, blue-shifted DW generation with high conversion efficiency at around 200 nm has been demonstrated by exciting LP$_{02}$-like mode in an uniform Ar-filled HC-ARF with experimentally feasible fiber and pulse parameters. The generated DW has an efficiency of $>$35\% when the fiber is filled under 10 bar Ar, pumped  in the anomalous dispersion regime of 1030 nm, 30 fs pulse duration, and 7 $\mu$J pulse energy. The high efficiency of the generated DW wave is due to the combination of self-phase modulation and anomalous dispersion, extreme shock effect and plasma-induced blue-shifted soliton. The temporal pulse is compressed down to less than single cycle of $\sim$1.5 fs. In comparison, the the nonlinear dynamics of LP$_{01}$-like mode is also investigated. When a LP$_{01}$-like mode is excited, the DW emits at a longer wavelength of $\sim$270 nm with efficiency of 15\%. However, the compressed pulse is not sufficient to ionize the gas due to the lack of sufficient free electron generation. 

% Section II
\section{Fiber modeling and Modal dispersion}
\begin{figure}
  \begin{center}
  \includegraphics[width=3.4in]{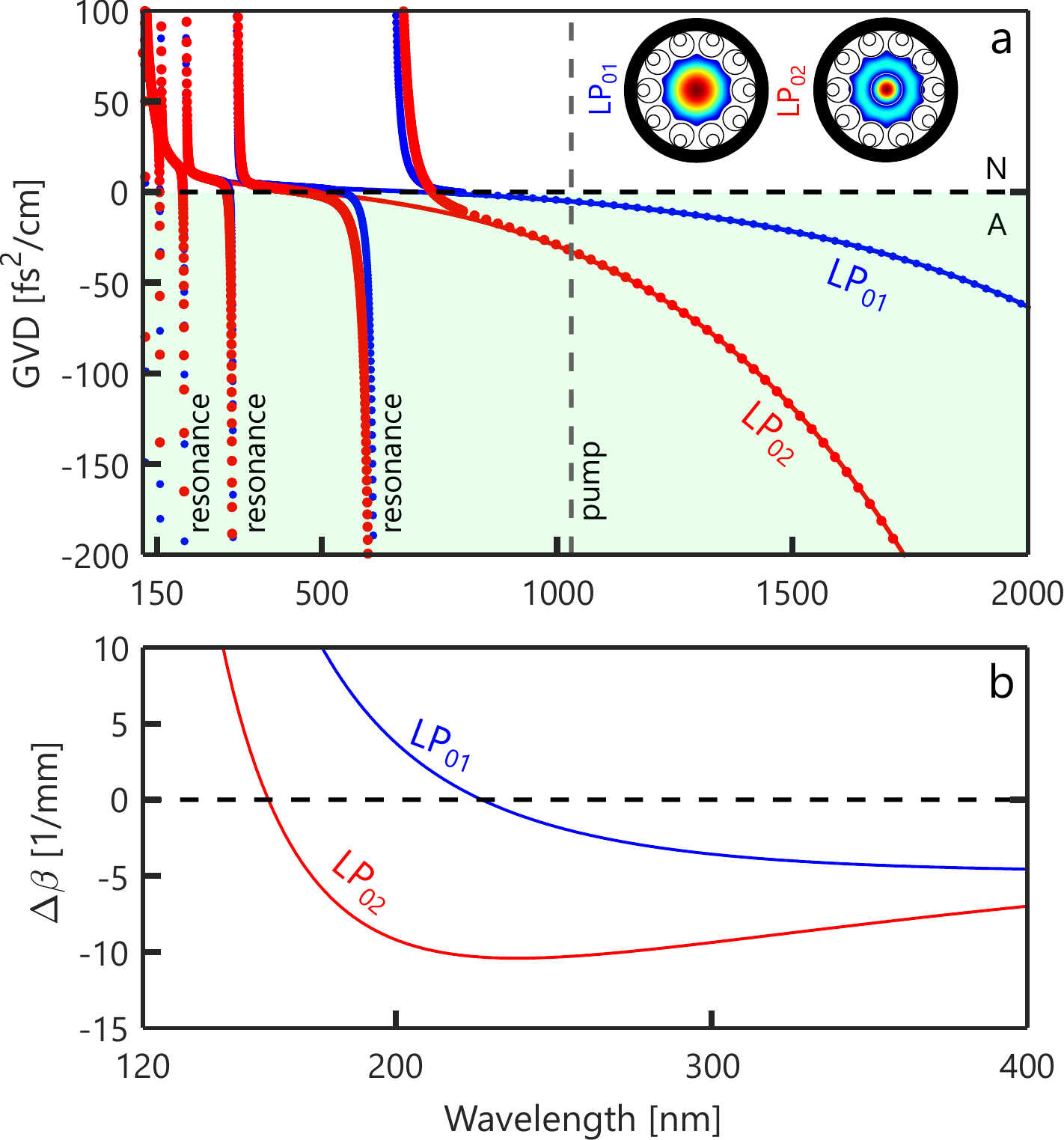}\\
   \caption{Numerically calculated (a) GVD of LP$_{01}$ (blue) and LP$_{02}$ (red)-like mode using modified capillary model \cite{hasan2018empirical} and FM modeling. solid: modified capillary model; dotted: FM modeling. The fiber has a core diameter of 40 $\mu$m, wall thickness of 300 nm, gap separation of 2 $\mu$m, and nested tube ratio of 0.5. The definition of fiber geometrical parameters can be found in \cite{habib2021impact}. Inset of (a): mode-field profile of LP$_{01}$ and LP$_{02}$-like mode. (b) propagation constant mismatch $\Delta\beta$ between soliton and DW according to \cite{travers2011ultrafast}. The contribution of ionization is not included. N: normal and A: anomalous dispersion. The light green shaded area presents the anomalous dispersion regime.}\label{fig:fig_1}
  \end{center}
\end{figure}

The HC-ARF considered in this work consists of 10-nested tubes with a core diameter of 40 $\mu$m, wall thickness of 300 nm, gap separation of 2 $\mu$m, and nested tube tube ratio, $d/D$ = 0.5. The wall thickness of 300 nm provides the center of first-order resonance at 630 nm calculated using \cite{poletti2014nested}: $\lambda_r = \frac{2t}{m} \sqrt{(n_\text{g}^2-1)}$, where $t$ is the wall thickness, $n_\text{g}$ is the refractive index of silica glass, and $m$ is the resonance order. The wall thickness is chosen 300 nm such that the resonance is far away from the laser pump wavelength of 1030 nm. The modal properties of the fiber was calculated using finite-element (FE) modeling based on \textsc{Comsol}$^{\circledR}$ software according to \cite{poletti2014nested,habib2019single,habib2021impact}. The HC-ARF can be designed multi-mode by carefully engineering the cladding parameters and number of tubes \cite{habib2019single}. The fiber is shown in  Fig.~\ref{fig:fig_1}(a) which can support more than 20 spatial modes and has maximum threshold loss of $<$100 dB/km. Fig.~\ref{fig:fig_1}(a) shows the mode-field profiles of LP$_{01}$ and LP$_{02}$-like modes which are well confined in the core and the loss of these modes are $<$10 dB/km at the pump wavelength of 1030 nm.

% figure-1

The propagation constant of LP$_{mn}$-like mode of HC-ARF can be represented using a capillary model \cite{marcatili1964hollow}:

\begin{equation}
\begin{split}
\beta_{mn}(\omega) = \sqrt{\frac{\omega^2n_\text{gas}^2(\omega,p)}{c^2}-\frac{U_{mn}^2}{R_\text{eff}^2}},
\end{split}
\label{eq:eq1}
\end{equation}

where $\omega$ is the angular frequency, $c$ is the velocity of light in vacuum, $n_\text{gas}(\omega,p)$ is the frequency and pressure dependent refractive index of the filling gas calculated using \cite{borzsonyi2008dispersion}, $U_{mn}$ is the $n$th zero of the $m$th-order Bessel function of the first kind (LP$_{01}$-like mode: $U_{01} = 2.450$ and LP$_{02}$-like mode: $U_{02} = 5.520$), $R_\text{eff}$ is the effective core radius according to \cite{hasan2018empirical}. The group velocity dispersion (GVD) of a 40 $\mu$m HC-ARF under 10 bar Ar was calculated using the capillary model which has been extensively used to study the ultrafast nonlinear dynamics in gas-filled fiber systems \cite{chang2011influence,chang2013combined,novoa2015photoionization,kottig2017mid}. In comparison, the GVD was also calculated using FE modeling and it is seen from Fig.~\ref{fig:fig_1}(a) that both models agrees very well. The only difference between the two models is that FE modeling can track the resonance bands which are determined by the fiber wall thickness. The zero-dispersion wavelength of LP$_{01}$ and LP$_{02}$-like modes are 712 nm and 481 nm respectively. At the pump wavelength, the calculated GVD of LP$_{01}$ and LP$_{02}$-like modes are --4.95 fs$^2$/cm and --32.91 fs$^2$/cm respectively. The phase mismatch ($\Delta\beta$) between the soliton and dispersive wave (DW) of LP$_{01}$ and LP$_{02}$-like modes are shown in Fig.~\ref{fig:fig_1}(b). The phase mismatch was calculated according to \cite{travers2011ultrafast} neglecting the ionization effect. It can be seen from Fig.~\ref{fig:fig_1}(b) that the DW of a pure LP$_{02}$-like mode is blue shifted compared to the LP$_{01}$-like mode since the zero-dispersion wavelength of LP$_{02}$-like mode resides in the blue side.

\section{Pulse propagation equation}

The optical pulse propagation was studied using a unidirectional pulse propagation equation which includes free-electron effects but without considering polarization and coupling between the different modes can be expressed as \cite{chang2011influence,chang2013combined,novoa2015photoionization}:

\begin{equation}
\begin{split}
\partial_{z}E(z,\omega) =  i\beta_{mn}(\omega-\frac{\omega}{v_g})E(z,\omega)+i\frac{\omega^2}{\mu_0}\mathcal{F}\bigg\{{\epsilon_0\chi^{(3)}E(z,t)^3}\bigg\} \\ 
-\frac{\omega\epsilon_0}{2\beta(\omega)}\mathcal{F}\bigg\{\partial_t\rho(z,t)\frac{I_\text{p}}{E(z,t)}+\frac{e^2}{m_e}\int_{-\infty}^{t}\rho(z,t')E(z,t')\bigg\},
\end{split}
\label{eq:eq2}
\end{equation}

where $z$ is the propagation direction, $t$ is the time in the reference frame moving with the pump group velocity $v_g$, $E(z,\omega)$ is the electric field in the frequency domain, $\beta_{mn}(\omega)$ is the propagation constant, $\epsilon_0$ and $\mu_0$ are the permitivity and permeability of free space respectively, $\chi^{(3)}$ is the third order susceptibility, $\rho$ is the time varying free electron (plasma) density, $I_\text{p}$ is the first ionization energy, $e$ and $m_e$ are the charge and mass of electron respectively, and $\mathcal{F}$ denotes the Fourier transform. The
Raman contribution of silica was neglected due to the very low light–glass overlap ($<<$1\%) and we assume nonradial dependence of plasma.  A quasi-static tunneling based on Ammosov, Delone, and Krainov (ADK) model was used to calculate the plasma density \cite{ammosov1987tunnel}.

% figure-2
\begin{figure}
  \begin{center}
  \includegraphics[width=3.4in]{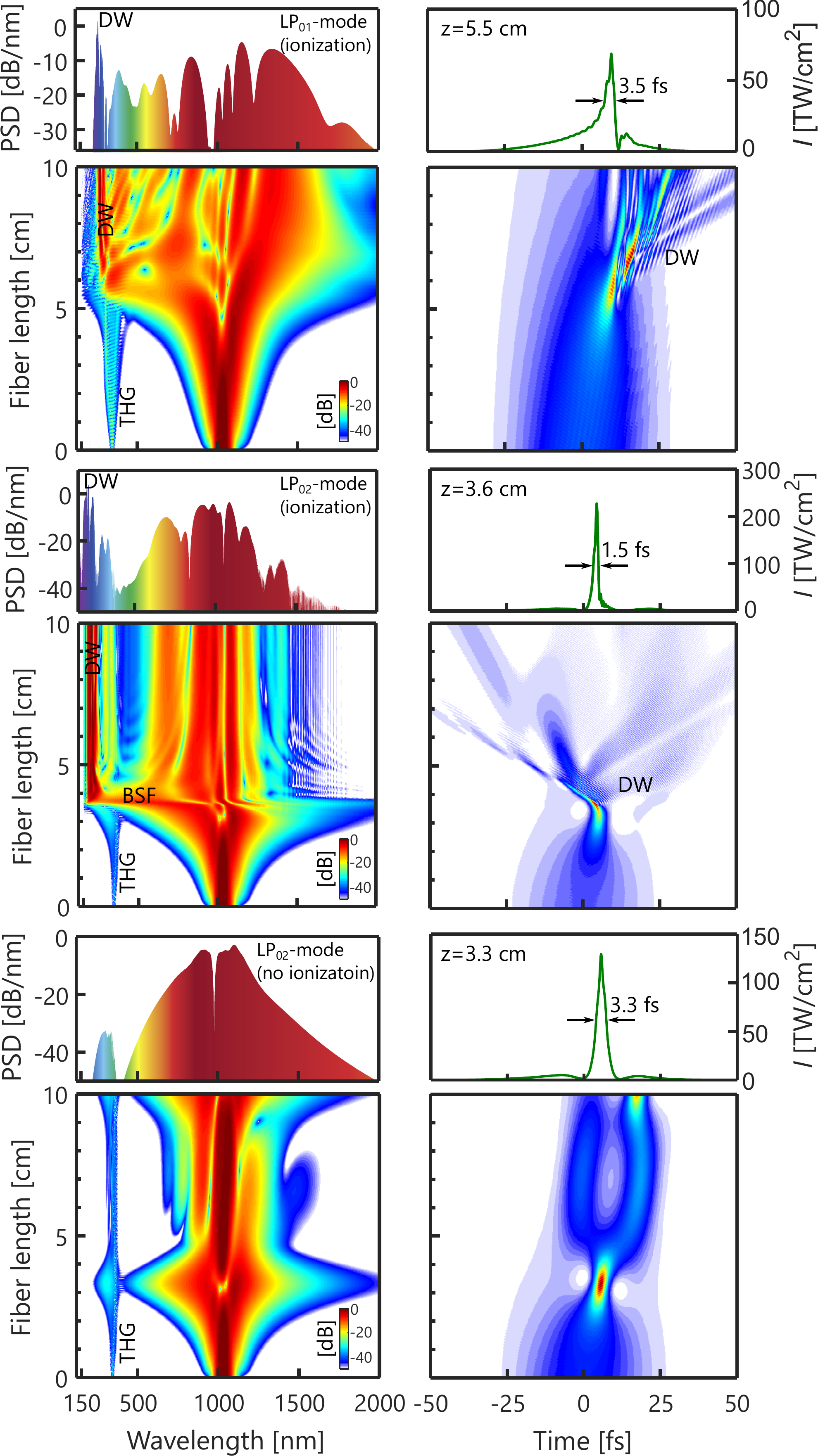}\\
    \caption{Numerically simulated spectral (left) and temporal (right) evolution for a 40 $\mu$m core HC-ARF under 10 bar of Ar, pule energy of 7 $\mu$J, pulse duration of 30 fs, and 1030 nm pump wavelength.  Spectral and temporal evolution of LP$_{01}$-like mode with ionization (top row), LP$_{02}$-like mode with ionization (middle row), LP$_{02}$-like mode without ionization (bottom row). The fiber has a wall thickness of 300 nm, gap separation of 2 $\mu$m, and nested tube ratio of 0.5. BSF: blue-shifted soliton, THG: third-harmonic generation. The simulations were performed using modified capillary model \cite{hasan2018empirical}. The PSD was calculated using: 
    $\text{PSD}=\frac{c}{\lambda^2}|E(z,\lambda)|^2f_\text{rep}$, where $f_\text{rep}$ is the laser repetition rate.
    }\label{fig:fig_2}
  \end{center}
\end{figure} 

\section{Numerical results and discussions}

\begin{figure}
  \begin{center}
  \includegraphics[width=3.4in]{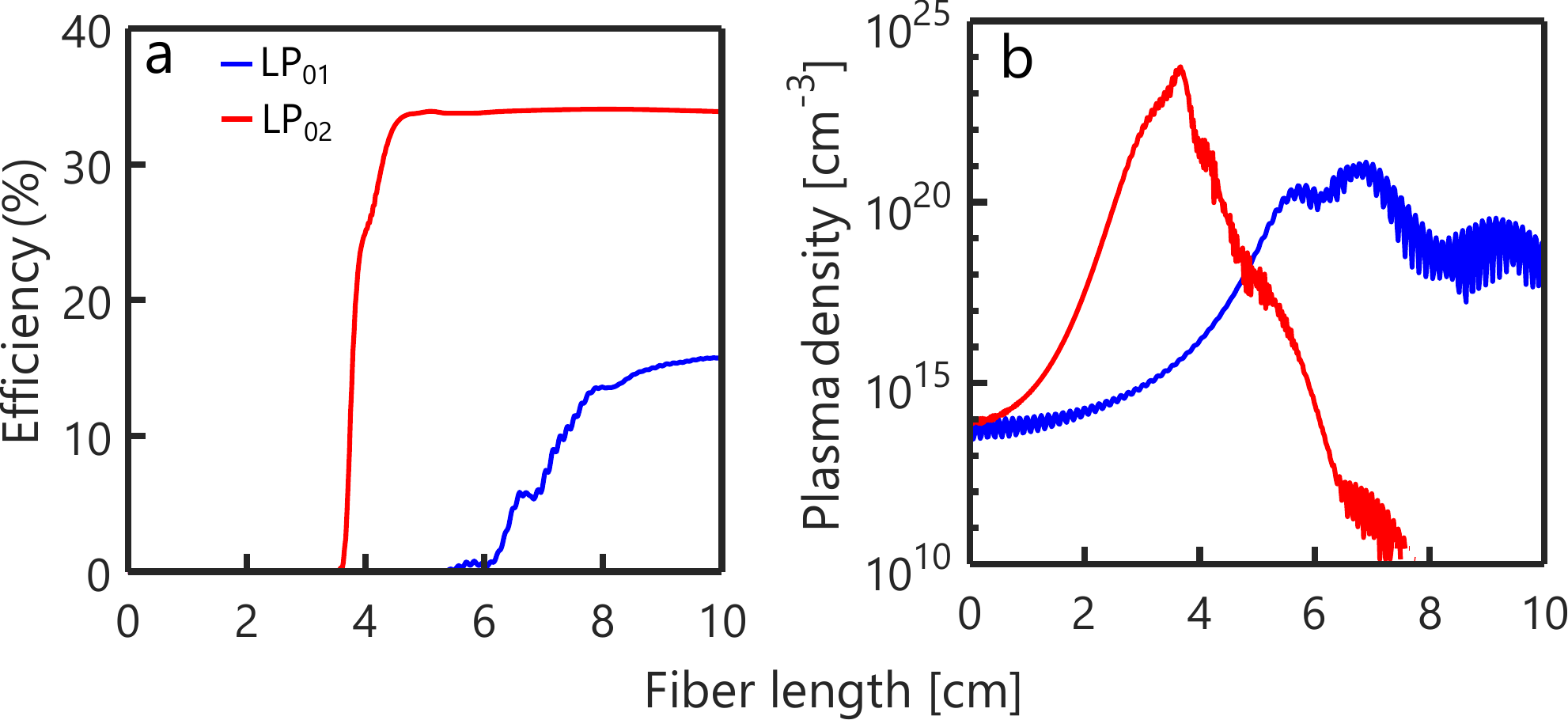}\\
  \caption{Efficiency (a) and average plasma density along the fiber propagation distance for LP$_{01}$ (blue) and LP$_{02}$ (red)-like mode. The maximum efficiency of LP$_{01}$ is $>$35\% whereas for LP$_{02}$-like mode is $\sim$15 \%. The plasma density was calculated using \cite{chang2013combined,ammosov1987tunnel}.}\label{fig:fig_4}
  \end{center}
\end{figure}

The spectral and temporal evolution of LP$_{01}$ and LP$_{02}$-like mode in a 10 cm long fiber filled with 10 bar Ar, pumped at 1030 nm with 30 fs pulse and pulse energy of 7 $\mu$J is shown in Fig.~\ref{fig:fig_2}. We begin our analyses by considering the excitation of LP$_{01}$-like mode. For fundamental LP$_{01}$-like mode, the pulse undergoes soliton self-compression down to $\sim$3.5 fs with peak intensity $>$65 TW/cm$^2$ after propagating $\sim$5.5 cm due to the interplay between self-phase modulation and anomalous dispersion. At the maximum compression point, a DW is emitted at $\sim$270 nm in the normal dispersion regime and a supercontinuum generation (SC) spanning 270--1800 nm. However, the peak intensity of the compressed pulse was not enough to ionize the gas since the plasma density at the maximum temporal compression point was found around 10$^{14}$ cm$^{-3}$ (see Fig.~\ref{fig:fig_4}(b)). Therefore, a similar spectral and temporal evolution can be seen when the ionization effect is turned-off which is not shown in Fig.~\ref{fig:fig_2}. 

For LP$_{02}$-like mode, the pulse undergoes a strong soliton self-compression down to $\sim$1.5 fs (less than a single cycle) with peak intensity $>$220 TW/cm$^2$ at an early stage of $\sim$3.6 cm, and a strong DW is emitted at 200 nm. Such a high intense pulse is enough to ionize the gas and form a plasma and a blue-shifted soliton (BSF) appears in the spectral profile. Unlike the SC of LP$_{01}$-like mode, the SC of LP$_{02}$-like mode is narrower at the longer wavelengths that can be seen after the temporal compression which is due to the high ionization loss \cite{saleh2011theory}. The spectral and temporal evolution of LP$_{02}$-like mode without ionization is shown in Fig.~\ref{fig:fig_2} (bottom row). When the ionization effect was not considered, DW is not generated in the UV regime since the energy of the compressed pulse is not sufficient to phase match to the DW. It turns out that the ionization effect is responsible for the generation of DW for LP$_{02}$-like mode and plays an important role of DW generation.

\begin{figure}
  \begin{center}
  \includegraphics[width=3.4in]{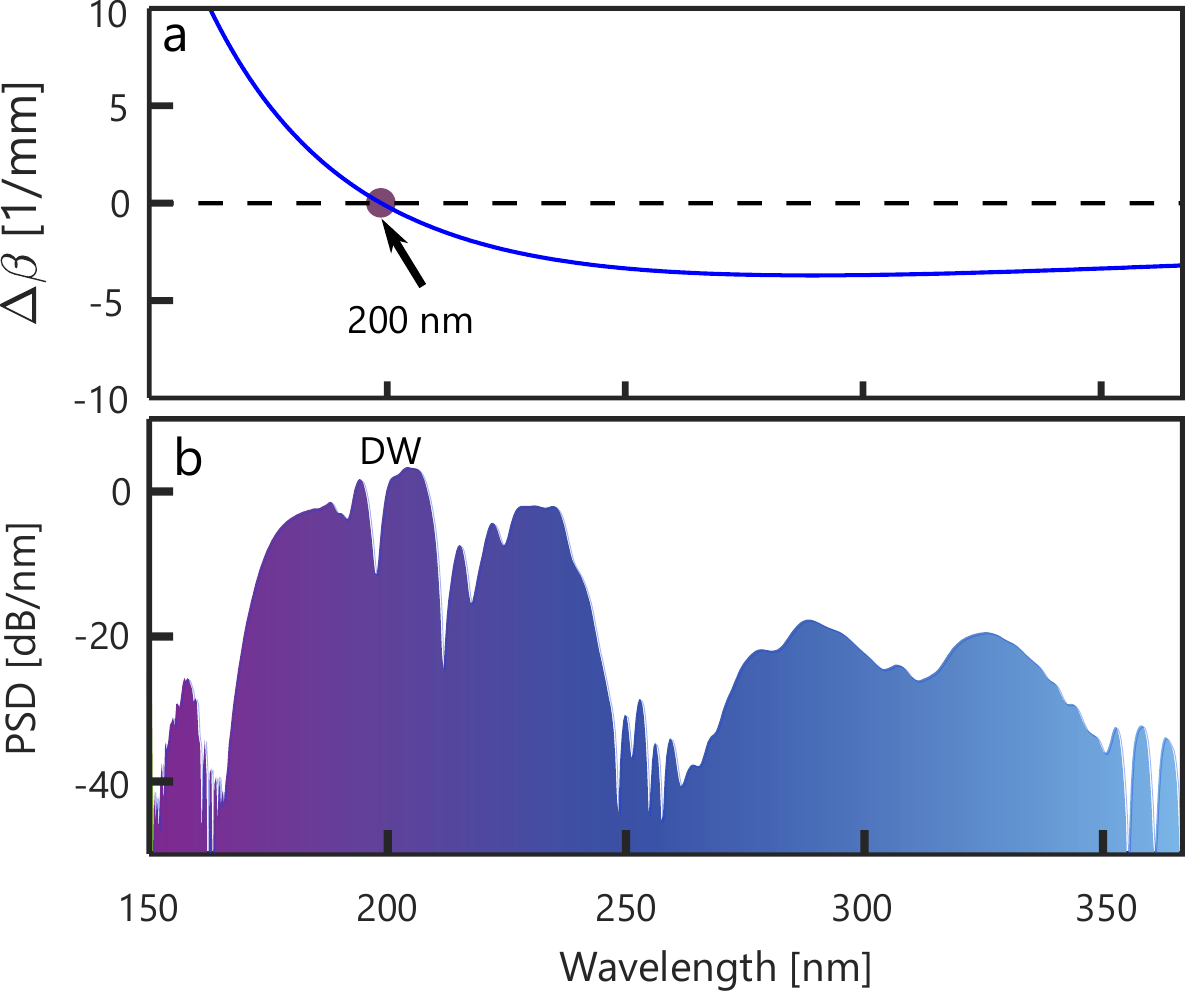}\\
  \caption{(a) Phase mismatch ($\Delta\beta$) between a soliton and DW for LP$_{02}$-like mode according to \cite{kottig2017mid}. The 40 $\mu$m HC-ARF was filled with 10 bar Ar, pumped at 1030 nm with a pulse energy of 7 $\mu$J and 30 fs pulse duration. The ionization was included to calculate the phase mismatch. The plasma density of $>$ 10$^{18}$ cm$^{-3}$ is calculated at maximum temporal compression. (b) The normalized PSD with UV dispersive wave emission.}\label{fig:fig_3}
  \end{center}
\end{figure}

To confirm the DW emission of LP$_{02}$-like mode at 200 nm, the phase mismatch curve ($\Delta\beta$) between the soliton and the DW including the ionization effect is shown in Fig.~\ref{fig:fig_3}(a). The phase mismatch can be expressed as \cite{kottig2017mid}:

\begin{equation}
\begin{split}
\Delta\beta = \beta(\omega)-\bigg(\beta_0+\frac{1}{v_g}[\omega-\omega_0]+\gamma P_p\frac{\omega}{\omega_0}-\frac{\omega_0}{2n_0c}\frac{\rho}{\rho_{cr}}\frac{\omega_0}{\omega}\bigg),
\end{split}
\label{eq:eq3}
\end{equation}

where $\beta_0$ is the propagation constant at the soliton centre frequency $\omega_0$, $\gamma$ is the nonlinear fiber parameter at the pump wavelength \cite{travers2011ultrafast}, $P_p$ is the soliton peak power of the compressed pulse, $n_0$ is the linear refractive  index of the medium, $\rho_{cr} \equiv \epsilon_0 m_e \frac{\omega_0^2}{e^2}$ is the critical plasma density \cite{couairon2007femtosecond}. It can be seen from Fig.~\ref{fig:fig_3}(a) that the phase matching occurs at 200 nm which agrees well with the DW wave emission seen from the power spectral density (PSD) curve of Fig.~\ref{fig:fig_3}(b). It should be noted that when the ionization effect is considered the phase matching condition occurred at longer wavelengths since the pump soliton shifts toward the blue side \cite{saleh2011theory} (see Fig.~\ref{fig:fig_2} (middle row)).  

% figure-3

% figure-4

The DW efficiency and plasma density of LP$_{01}$ and LP$_{02}$-like modes are shown in Fig.~\ref{fig:fig_4}(a-b). The DW efficiency of LP$_{02}$-like mode is $>$35\% whereas the efficiency of LP$_{01}$-like mode is around 15\%.  The dramatic enhancement of the DW emission for LP$_{02}$-like mode is
due to the strong shock effect and ionization-induced blue-shift\cite{saleh2011theory}. The plasma density of LP$_{02}$-like mode is above $>$10$^{18}$ cm$^{-3}$ at the maximum temporal compression point which is enough to ionize the gas. The free electron generation decays drastically after the temporal compression as it propagates through the fiber due to the high ionization loss. However, at the temporal compression point, the plasma density of LP$_{01}$-like mode is only  10$^{14}$ cm$^{-3}$.

% figure-5
\begin{figure}
  \begin{center}
  \includegraphics[width=3.4in]{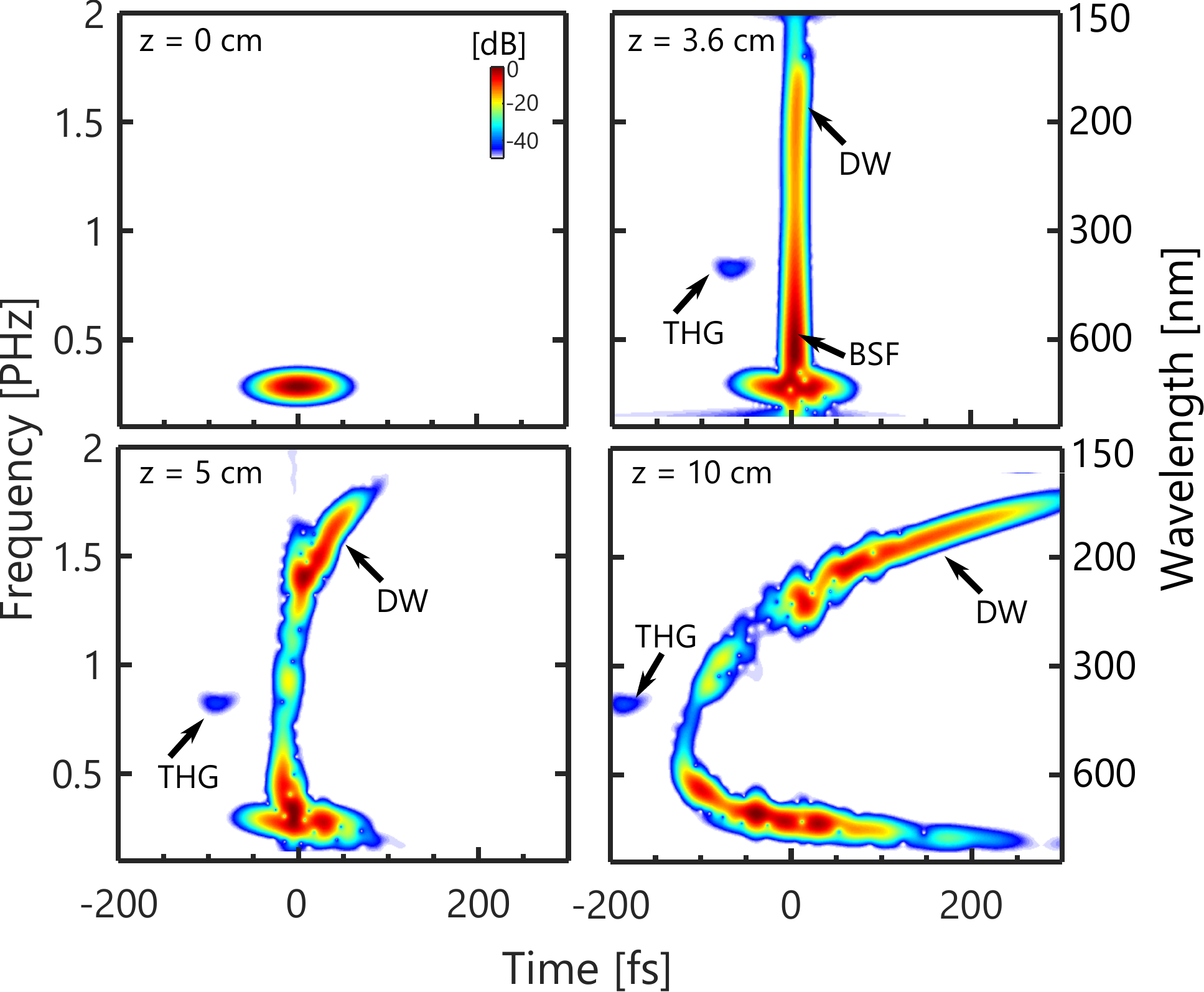}\\
  \caption{ Calculated spectrogram of LP$_{02}$-like mode for a 40 $\mu$m core HC-ARF under 10 bar of Ar, pule energy 7 $\mu$J, pulse duration 30 fs, and 1030 nm pump wavelength at selected distances. The ionization is included in the simulation. The spectrogram was calculated using a 10 fs Gaussian gate pulse. The third harmonic of the pump pulse is evident at 343.33 nm.}\label{fig:fig_5}
  \end{center}
\end{figure}

To get more physical insights on the underlying mechanism of DW emission for the LP$_{02}$-like mode, the spectrograms are plotted in Fig.~\ref{fig:fig_5} using the cross correlation frequency-resolved optical gating for different selected fiber lengths. A 10 fs Gaussian pulse were used to plot these spectrograms. The location of DW, blue-shifted soliton (BSF), third-harmonic generation (THG) are shown. At the propagation distance of $z=3.6$ cm, the pulse undergoes a maximum temporal compression and the spectrum extends towards the blue due to the plasma formation and a DW is emitted in the normal dispersion regime at around 200 nm. During the propagation, the DWs broadens linearly in time which can be seen for $z=$ 5 cm and $z=$10 cm. 

% Conclusion
\section{Conclusion}
In conclusion, a new approach to enhance the efficiency of a DW generated in the UV regime by exciting LP$_{02}$-like mode in Ar-filled HC-ARF has been demonstrated. The numerical modelings are based on the experimentally feasible fiber and pulse parameters that can predict DW generation at 200 nm followed by SC spanning 200--1500 nm. The formation of extreme shock effect and plasma induced blue-shifted soliton leads to an efficiency of $>$35\%. The numerical modeling predicts that it requires only  3.6 cm of fiber to generate the DW. The DW can be further blue-shifted if higher LP$_{0n}$ are excited but requires more pump energy to phase match the soliton and DW. The results presented in this work may be valuable towards the development of compact, bright, and highly efficient UV light source.

%\section*{Acknowledgment}

%The author would like to thank Dr. Mohammed Saleh for useful discussions.

\bibliographystyle{IEEEtran}
\bibliography{IEEEabrv,Bibliography}
\vfill
%\balance
\end{document}